\documentclass{emulateapj}

\shorttitle{On the CMRD and IMF}
\shortauthors{Reggiani et al.}

\begin{document}

\title{Binary Formation Mechanisms: Constraints from the Companion Mass Ratio Distribution}

\author{Maddalena M. Reggiani\altaffilmark{1}}
\author{Michael R. Meyer\altaffilmark{1}}

\altaffiltext{1}{Institute of Astronomy, ETH Zurich, CH-8093 Zurich, Switzerland; reggiani@phys.ethz.ch}

\begin{abstract}
We present a statistical comparison of the mass ratio distribution of companions,
as observed in different multiplicity surveys, to the most recent estimate of the single
object mass function (Bochanski et al. 2010). The main goal of our analysis is to
test whether or not the observed companion mass ratio distribution (CMRD) as a
function of primary star mass and star formation environment is consistent with
having been drawn from the field star IMF.
We consider samples of companions for M dwarfs, solar type and intermediate mass stars, both in the field as well as clusters or associations, and  compare them with populations of binaries generated by random pairing from the
assumed IMF for a fixed primary mass.
With regard to the field we can reject the hypothesis that the CMRD was drawn from the IMF for different primary mass ranges: the observed CMRDs show a larger number of equal-mass systems than predicted by the IMF. This is in agreement with fragmentation theories of binary formation. For the open clusters $\alpha$ Persei and the Pleiades we also reject the IMF random-pairing hypothesis.
Concerning young star-forming regions, currently we can rule out a connection between the CMRD and the field IMF in Taurus but not in Chamaeleon I. Larger and different samples are needed to better constrain the result as a function of the environment. 
We also consider other companion mass functions (CMF) and we compare them with observations. Moreover the CMRD both in the field and clusters or associations appears to be independent of separation in the range covered by the observations. Combining therefore the CMRDs of M (1-2400 AU) and G (28-1590 AU) primaries in the field and intermediate mass primary binaries in Sco OB2 (29-1612 AU) for mass ratios, $q=M_2/M_1$, from 0.2 to 1, we find that the best chi-square fit follows a power law $dN/dq \propto q^{\beta}$, with $\beta=-0.50\pm0.29$, consistent with previous results. Finally we note that the KS test gives a $\sim$1\% probability of the observed CMRD in the Pleiades and Taurus being consistent with that observed for solar type primaries in the field over comparable primary mass range. This highlights the value of using CMRDs to understand which star formation events contribute most to the field.
\end{abstract}


\section{Introduction}
The study of binary stellar systems and their properties is one of the most important topics in star formation.
Since many stars form in multiple systems, both in the field \citep[e.g][]{Duquennoy1991,Fischer1992} and in star-forming regions \citep[e.g][]{Patience2002}, a correct understanding of the binary fraction and
companion mass distribution for different primary masses represents a fundamental test for star formation theories.
Binary populations, in fact, may carry an even wider amount of information concerning star formation processes than the IMF \citep{Goodwin2009}.
By conventional definition, in a binary system of stars with masses $M_{1}$ and $M_{2}$, with $M_{1}>M_{2}$, $M_{1}$
and $M_{2}$ indicate the primary and secondary mass, respectively. Consequently one can define the ratio of
the secondary over the primary mass as $q=M_{2}/M_{1}$.
Analogous to the initial mass function (IMF) for single objects, one can define the companion mass ratio
distribution (hereafter CMRD) as the distribution of $q$ for a chosen primary mass.

Different binary formation models predict different mass ratio distributions and dependencies of the CMRD on the primary mass.
Traditionally these classes of models have been divided into capture and fragmentation scenarios.
Capture refers to the tidal capture of two unbound objects on a timescale that is long compared to the collapse time of each component \citep[e.g.][]{McDonald1993}. For each primary star the mass of the secondary is chosen randomly from the single star mass function and the secondary-mass distribution would reflect the IMF. While tidal capture appears to be too inefficient in reproducing high binary fractions, it has been noticed that, particularly in small groups of stars, star-disk encounters may form binaries \citep{McDonald1995}. In any case even this disk assisted capture, whereby a star passing through the disk of another which dissipates enough kinetic energy to form a bound system, is unlikely to be the most relevant binary formation mechanism \citep{Boffin1998}.

Fragmentation scenarios are the preferred mechanism for the formation of multiple systems. The so-called fragmentation models are usually classified as prompt fragmentation \citep[e.g.][]{Boss1986,Bonnell1992} and disk fragmentation \citep[e.g.][]{Bonnell1994,Stamatellos2009}. In the prompt fragmentation scenario both primary and secondary stars form by fragmentation of the same collapsing molecular cloud core. Disk fragmentation takes place in a newly formed star-disk system in which the disk subsequently fragments due to density perturbations. The latter mechanism is a process by which low-mass stars and BDs (e.g. companions with $q \le 0.25$) may form.
In both cases fragmentation is only the first step in binary formation and processes such as disk accretion and dynamical interactions all contribute to determine the final properties of binary systems \citep{Bate2004}. In general, continued accretion onto both objects from a common reservoir tends in the long term to equalize the masses, moving the $q$ distribution towards unity, and this effect seems to be more significant for high mass primaries and in closer binaries \citep{Bate2000}.

Recently, a variation on capture has been proposed as mechanism for forming wide binaries \citep{Kouwenhoven2010, Moeckel2010}. In this scenario wide binaries ($10^4$ - $10^5$ AU)  would form during the dissolution phase of star clusters, especially during the quick expansion of clusters after gas expulsion. This mechanism could perhaps explain the substantial population of wide binaries observed in the field and, to the first order approximation, the mass ratio distribution could be similar to that expected from random pairing of individual stars \citep{Kouwenhoven2010}. 

However, at the moment no model is able to reproduce all of the observed binary properties, in particular, the predicted distributions of separations and mass ratios tend not to match the observations very well \citep{Goodwin2007}.
Even if no observation will definitely confirm one theory, observations of CMRDs in disparate environments can at least put constraints on theoretical binary formation models.
Beginning with the pioneering multiplicity survey of G stars in the solar neighborhood by
\citet{Duquennoy1991}, there have been several studies of binary properties in the field
(e.g. \citet{Fischer1992}, \citet{Reid1997} for M dwarfs, \citet{Metchev2009} and  \citet{Raghavan2010} for solar-type stars) and in clusters or associations \citep[e.g.][]{Leinert1993,Ghez1993,Patience2002,Kraus2011}. Curiously  in young clusters the binary fractions overall are higher but little is known about the CMRD.

Furthermore observations of different star-forming regions revealed that the mass distribution of cores usually has a form similar to the IMF \citep{Motte1998,Motte2001,Alves2007}, leading to the suggestion that IMF and the core mass function are directly related.
At the same time the majority of observations of single objects from the field, local young clusters, old globular clusters and associations suggest a universal IMF \citep{Bastian2010}.
What is the role played by binaries and multiple systems? Is there a connection between the CMRD and the IMF?
\citet{Metchev2009} present an analytical form of the companion mass function (CMF) and
reject the hypothesis that the CMRD of solar-mass stars in the field is consistent with having been drawn from random pairing of the single star IMF.
However, the most recent estimate of the log-normal IMF for isolated objects \citep{Bochanski2010} is peaked at higher masses than those they considered, and might be in closer agreement with the CMRD they derived. It is also interesting to study CMRDs as a function of primary mass as well as in different environments and look for variations in the CMRD among regions where the IMF varies.

Finally, another crucial question that the CMRD could address is the origin of the field. Matching properties of the field to 
star-forming regions could give insights into what sort of regions contribute most to the field. Binaries are subject to dynamical evolution and disruption \citep{Parker2009}, changing the overall binary fraction. The extent of these dynamics depends on the environment in which they were born. Recent results from N-body simulations also predict that some binary properties (e.g. the CMRD for low-mass binaries) might be independent of dynamical processing \citep{Parker2010}, making them excellent tracers of origins.

A rigorous approach to answering these questions requires a careful account of the completeness of observational data
and potential biases as a function of separation, besides a proper choice of the IMF.
\textit{So far, a complete analysis of the CMRD over a broad range of primary masses and as a function of separation and environment has not been done.}
In this paper we address the problem of the connection between IMF and CMRD by considering samples of binaries with
primaries of different masses and using the most recent evaluation of the IMF. We first describe the datasets we have used in our analysis (Section \ref{Datasets}). Then in Section \ref{Method}
we discuss the methodology we adopt while
the results obtained are shown in Section \ref{Results}. Finally, Section \ref{Discussion} and Section \ref{Conclusions} are left to the discussion of the  results and to our conclusions.

\section{Datasets} \label{Datasets}
\subsection{Samples from the field and Sco OB2}\label{Datasets_field}
As mentioned before, the multiplicity of stars in the field has been investigated in the past years by different groups. Among these surveys we have selected for our analysis three studies, each of which surveyed companions for a restricted range of primary masses (M dwarfs, solar-type stars and intermediate mass stars). In this section we give a brief overview of the datasets we have chosen while in section \ref{Cluster&associations} we describe the samples of binary systems from young clusters/associations that we have also considered in our analysis. A full summary of the main properties of all these samples is given in Table~\ref{table:Samples}. We already said that to obtain reliable results it is important to account for completeness and possible biases. However also the estimate of completeness and observational biases is not free from uncertainties. Therefore, instead of considering completeness-corrected samples, we decided to limit our investigation to the mass and separation range where the completeness of the samples is flat (so the sample is \textit{representative} in mass, if not complete) and exceeds a certain level ($\ge 65$\%).

\begin{deluxetable*}{cccccc}
\tablewidth{0pt}
\tablecaption{Sample Properties \label{table:Samples}}
\tablehead{\colhead{Sample} & \colhead{Ref}\tablenotemark{a}  & \colhead{Primary Type} & \colhead{No. Multiple Systems} & \colhead{Separation Range (AU)} & \colhead{q$_{lim}$}}
\startdata
Field & 1  & M  & 27 & 1-2400 & 0.2\\
Field & 2 &  F/G &  30 &  28-1590 & 0.1\\
ScoOB2 & 3 & A/late-B &  60 & 29-1612 & 0.05\\
Pleiades & 4 & F/G & 22 & 11-910 & 0.2\\
$\alpha$ Persei & 5 & F/G &  18 & 26-581 & 0.25\\
Chamaeleon I & 6 & G/K\tablenotemark{b} & 13 & 20-800 & 0.1\\
Taurus & 7 & G/K\tablenotemark{c} & 40 & 5-5000 & 0.1\\
\enddata
\tablenotetext{a}{References: (1) \citet{Fischer1992} , (2) \citet{Metchev2009}, (3) \citet{Kouwenhoven2005}, (4) \citet{Bouvier1997}, (5) \citet{Patience2002}, (6) \citet{Lafreniere2008}, (7) \citet{Kraus2011}.}
\tablenotetext{b}{The mass range is 0.55 and 2.2 M$_{\odot}$, comparable to MH09}
\tablenotetext{c}{The mass range is 0.7 and 2.7 M$_{\odot}$, comparable to MH09}
\end{deluxetable*}

The sample of M dwarfs we have considered is the set of binary systems collected by  \citet{Fischer1992} (hereafter FM92) from several high-quality surveys of M dwarfs with distances within 20 pc from the Sun (ages $\ge$ Gyr). Each one of these surveys covers a different angular separation range, but the complete sample extends from roughly 1 to 2400 AU in separation and down to $q=0.2$ in mass ratio.
Generally, M dwarfs with masses $<0.2$ M$_{\odot}$ show mass ratios biased towards unity due to sensitivity limitations \citep{Fischer1992}.
For this reason we have considered only binary systems with primaries having masses between 0.2 to 0.55 M$_{\odot}$, where the sample is 85\% complete. The sample consists then of 27 systems.

Regarding solar-mass stars we selected the work presented in \citet{Metchev2009} (hereafter MH09).
MH09 report results from an adaptive optics survey of stellar and substellar companions
to solar analogs (range in primary mass between 0.7 to 1.3 M$_{\odot}$) within 10-190 pc and in the 3 Myr - 3 Gyr age range.
The orbital separation interval covered is 28-1590 AU.
The choice of this survey, with respect to previous works  \citep[e.g.][]{Duquennoy1991}, is due to the
higher sensitivity to low-mass companions, meaning small mass ratios ($q\leq0.1$).
In order to have a 65\% complete sample we considered the set of 30 binary systems with $q\geq0.1$ and companions between  28 and 1590 AU from the primary which was defined as minimally-biased sample ($AD_{30}$) in their paper.

Finally, we chose a sample of companions to A-type and late B-type primaries. Due to the shorter lifetime of more massive stars and the difficulty to find a statistically large and complete survey of intermediate mass primaries in the field, we selected a dataset in the young (5-20 Myr) and nearby ($\sim$140 pc) Scorpius OB2 association (ScoOB2).
This binary population was observed in the near-infrared adaptive optics multiplicity survey described by \citet{Kouwenhoven2005} (hereafter K05) and the properties of the 60 stellar systems we have used in our analysis are taken from
\citet{Kouwenhoven2007}. This survey is sensitive to very low mass ratios (down to $q\sim 0.05$) over the orbital separations range 29-1612 AU between primaries and companions.
Despite the fact that Sco OB2 is not a sample from the field and it is young, it is still the best sample ($\ge 90\%$ complete) of intermediate mass primary binaries we can study. Therefore we will include it in the analysis of the other datasets from the field.

We emphasize that the three datasets span similar separation ranges. Even though the study of the detailed dependence of the CMRD on angular separation goes beyond the purpose of the present work, in section \ref{Results_field} we will show that our results should not be affected by any possible change in the shape of the CMRD with orbital separation.

\subsection{Clusters or associations}\label{Cluster&associations}
We considered also four datasets of companions to solar-type stars from nearby clusters or associations that over similar separation ranges have a reasonable number of binary systems.
We have selected observations of binaries in two open clusters, Pleiades and $\alpha$ Persei ($\alpha$ Per), and in two T associations, Chamaeleon I and Taurus.

The Pleiades is one of the best studied open clusters, due to its proximity and richness (roughly 1000 stars at a distance of $\sim$ 120 pc). With an age of 125-150 Myr \citep{Stauffer1998,Burke2004} it is just old enough to be dynamically evolved. The sample from the Pleiades \citep{Bouvier1997} consists of 22 binary systems with G and K primaries observed in the near-IR using adaptive optics. The separation range covered by this survey is 11-910 AU and the mass ratio distribution is more than 70\% complete down to 0.2 over this separation range.

$\alpha$ Per is a younger cluster, with an age of $\sim$90 Myr \citep{Stauffer1999}, at a distance of $\sim$ 190 pc \citep{Robichon1999}.
We selected a sample of 18 solar type stars within the dataset presented in \cite{Patience2002}. They were nearly complete in the separation range from 26 to 581 AU and were sensitive to mass ratios $q>0.25$.

Chamaeleon I, instead, is one of the nearest \citep[$\sim$170 pc,][]{Bertout1999} low-density young \citep[$\sim$1 Myr,][]{Luhman2004} star-forming regions. It consists of $\sim$230 stars and has a stellar density that is low compared to other young regions \citep{Luhman2008}. We have considered the results of a multiplicity survey presented by \citet{Lafreniere2008}. The primaries span the mass range from $\sim$ 0.1 to 3 M$_{\odot}$ and the separation range $\sim$20-800 AU. We have selected a subsample with only K and G primary binaries with masses between 0.55 and 2.2 M$_{\odot}$ and mass ratios down to $q\sim 0.1$ ($\sim 90\%$ complete), comparable to MH09 (13 systems in total).

Finally we selected a sample of solar-type primary binaries in Taurus from the almost complete sample by \cite{Kraus2011}. Taurus is another young (1 Myr) low-density star-forming region close to the Sun (d = 140 pc) with more than 300 PMS stars and brown dwarfs \citep{Kenyon2008}. We considered 40 systems with primary masses between 0.7 and 2.5 M$_{\odot}$, mass ratio $q\ge 0.1$ and angular separation in the range $5-5000$ AU. 

We will discuss the dependence of the CMRD on separation in section \ref{Results_field}.

\section{Methodology for our analysis of these surveys}\label{Method}
\subsection{Monte Carlo simulations} \label{MC sim}
Our goal is to explore whether the observed mass ratios in the field  and young clusters or associations as function of primary mass could be the outcome of random pairing of stars from the stellar IMF.
To this end we have created a Monte Carlo tool able to generate artificial companion mass ratio distributions as expected by random sampling of secondaries from a chosen function, for fixed primary mass.
Through these simulations one can reproduce a population of N binaries by fixing the mass of the primary and the analytic form to be tested, and then compare this simulated CMRD with the observations.

\subsection{Initial Mass Function}
The Initial Mass Function we have considered in our analysis is the single objects IMF from \citet{Bochanski2010} (hereafter Bo2010).
Below 1 $M_{\odot}$ it is a log-normal function of the mass, defined as $\xi(\log m)=d n / d \log m$.
Except for a normalization constant, it can be parametrized (in ($\log M_{\odot})^{-1} pc^{-3}$) as:
\begin{equation}
\xi(\log m)\propto \exp \left\{-\frac{(\log m - \log m_{c})^2}{2\sigma^2}\right\}
\end{equation}
where $m_c=0.18$ and $\sigma= 0.34$.

For m$>$1 M$_{\odot}$ we assumed the classical "Salpeter slope":
\begin{equation}
\xi(\log m)\propto m^{-1.35}.
\end{equation}

The study by \citet{Bochanski2010} is based on the observational work presented in \citet{Covey2008} which represents the largest field investigation of the luminosity function to date constructed from a catalog of matched Sloan Digital Sky Survey (SDSS) and Two Micron All Sky Survey (2MASS) sources.
Note that in Bo2010 the log-normal peak of the mass distribution is shifted toward higher masses compared to
\citet{Chabrier2003} ($m_c=0.08$ and $\sigma=0.69$).

In each run of our Monte Carlo simulations, the assumed IMF is normalized to the primary mass that we choose, to the appropriate range of $q$ for the dataset with which we compare the results, and to the number of binaries $N$ that we want to reproduce. We typically run each simulation $10^5$ times.

\subsection{KS test} \label{KS_test}
To evaluate the probability that the observed CMRDs and the simulated ones come from the same parent distribution we apply the Kolmogorov-Smirnov test (KS test) to the cumulative distributions of $q$ values. The KS test is a statistical test which returns the probability that two distributions were drawn from the same parent sample by examining the maximum difference in the cumulative distribution functions. 

For our purposes, we have tested how well this statistical tool can distinguish differences in the shape of the distributions and evaluated the extent to which results depend on sample size.

First, we tested the reliability of the KS test to distinguish two populations of stars distributed in mass according to power law distributions with different slopes as a function of sample size. With larger populations, the KS test is able to detect smaller differences in slope. From the comparison of distributions of only 10 objects the KS test gives a probability of $\sim 10^{-2}$ for a difference of 5 in the slopes while when the number of objects increases to 30 the same probability is already obtained with a difference of 2.5.

\begin{figure}
\includegraphics[scale=.35]{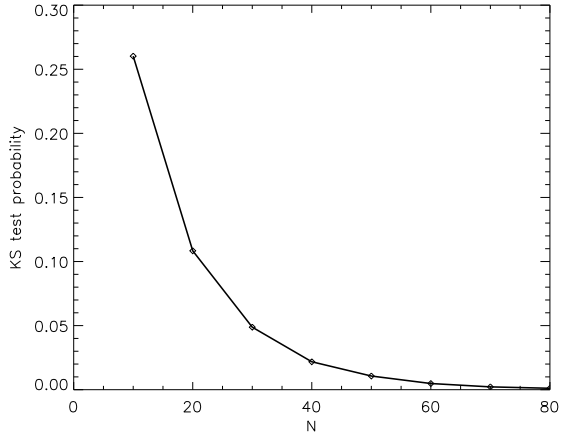}
\caption{\small{\textit{Capability of the KS test to distinguish flat and log-normal distribution.} The solid black line describes how varies the probability given by the KS that a log-normal differs from a flat distribution, as function of sample size. We computed this probability for samples of 10, 20, 30, 40, 50, 60, 70 and 80 objects. }}\label{test_diff_flat}
\end{figure}

Second, we checked the capability of the KS test to distinguish a log-normal and a flat distribution, again as function of sample size. This test is of great importance because, on one hand, we want to test the hypothesis of a log-normal CMRD, on the other, the linearly flat distribution of $q$ is a commonly made assumption in  numerical simulations \citep[e.g.][]{Kouwenhoven2009,Parker2009}. In Figure \ref{test_diff_flat} we show our results. In this comparison, with a sample of $\sim$50 objects the KS test returns a  1\% chance of having been drawn from the same parent. A KS probability of 1\% is the threshold we adopt equal to or below which we reject the hypothesis of two distributions being consistent.

\section{Results} \label{Results}
Here, we report our findings regarding the comparison of the observed CMRDs with the simulations. We begin describing in section \ref{Results_field} the results for the samples of M dwarfs, G stars in the field and intermediate mass stars in ScoOB2.  In section \ref{Results_young} we summarize the outcome of our tests for the Pleiades, $\alpha$ Per, Chamaeleon I and Taurus.
We compare the observed CMRDs with other commonly assumed companion mass functions in section \ref{Results_flat}. Finally (section \ref{Chi-square best fit}) we give our best-fit estimate of the combined distribution of M dwarfs and G stars in the field and intermediate mass stars in ScoOB2.

\subsection{Results from the field and Sco OB2} \label{Results_field}
\begin{figure}
\includegraphics[scale=.35]{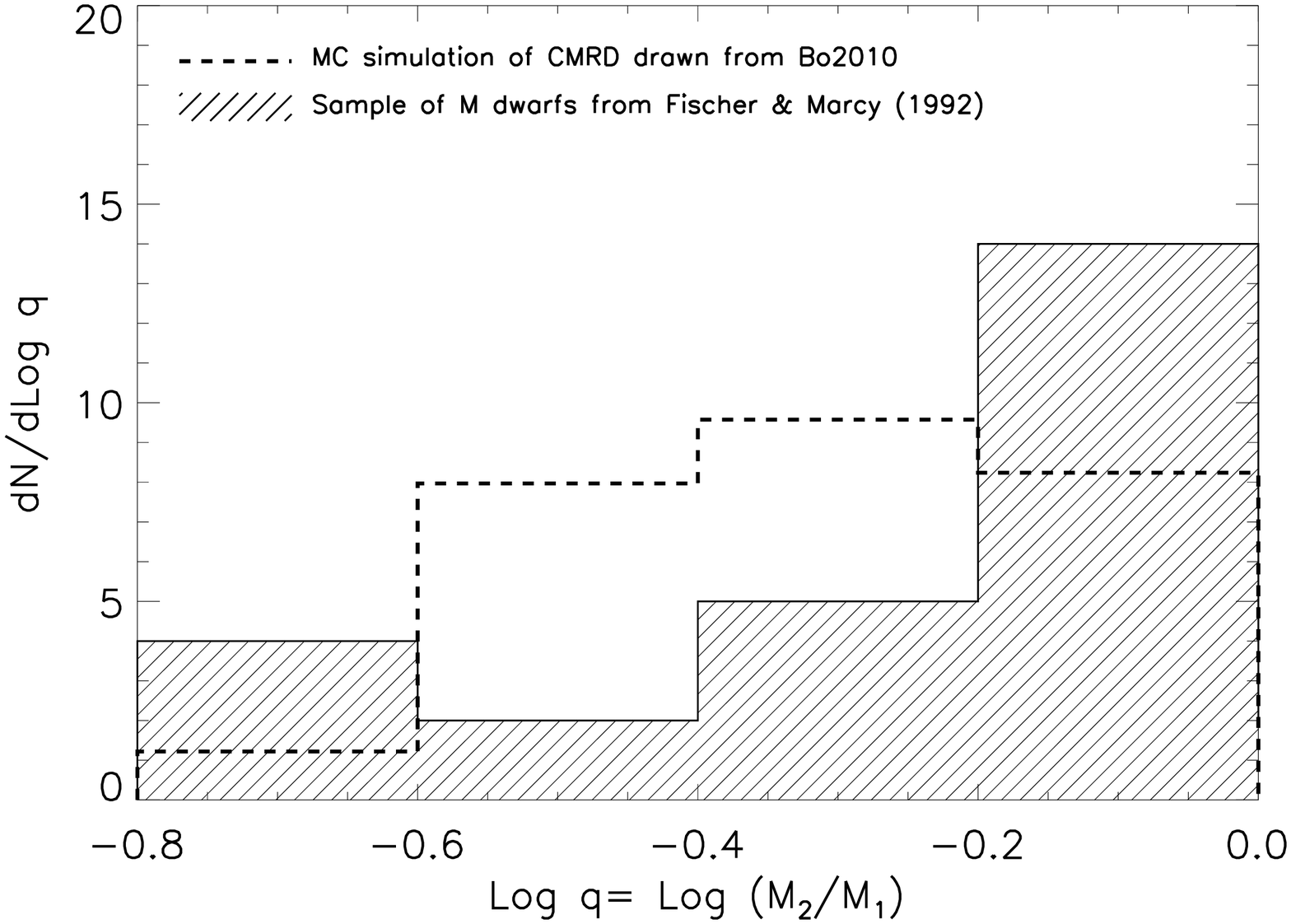}
\vspace{0.3cm}\\
\includegraphics[scale=.35]{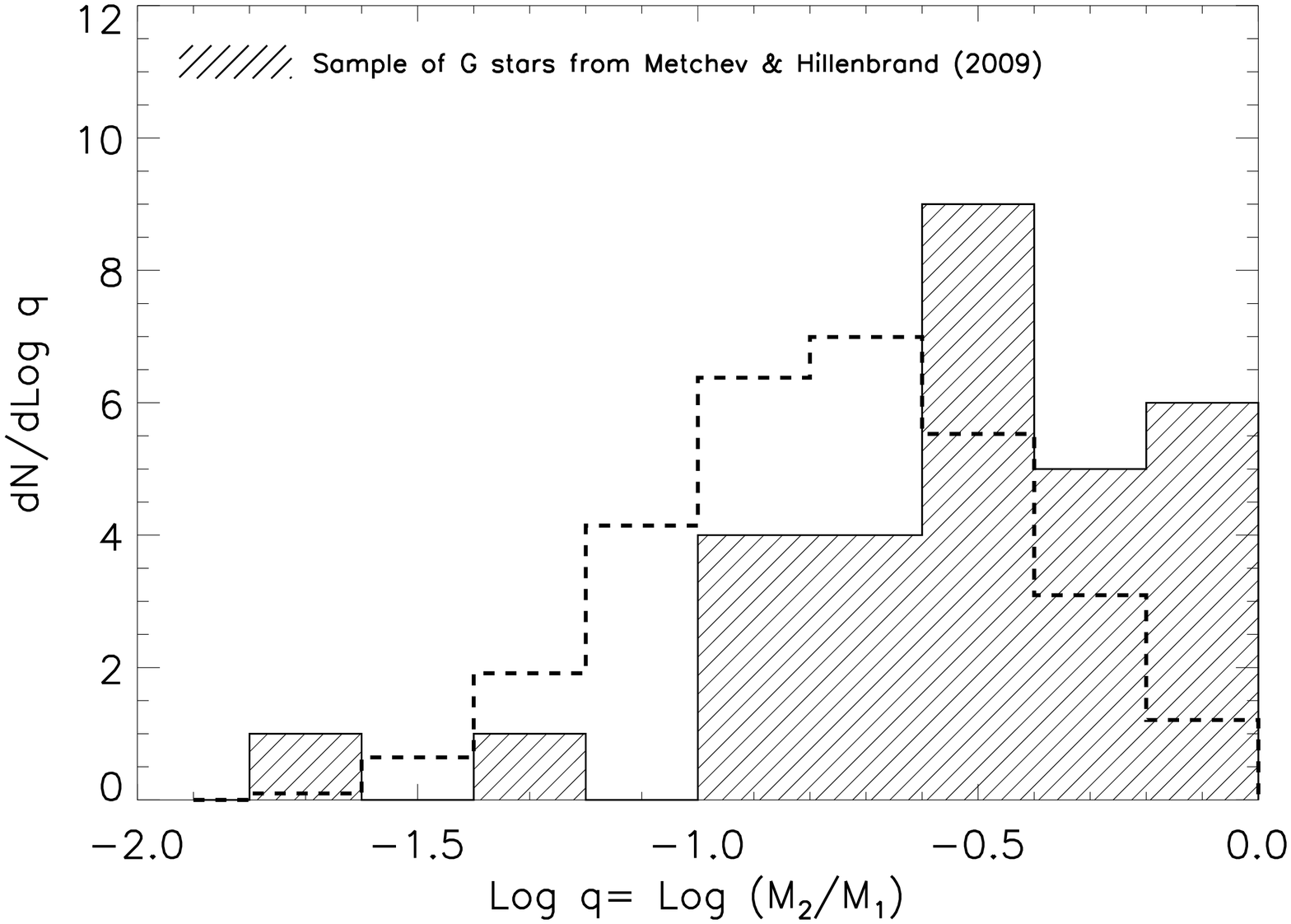}
\vspace{0.3cm}\\
\includegraphics[scale=.35]{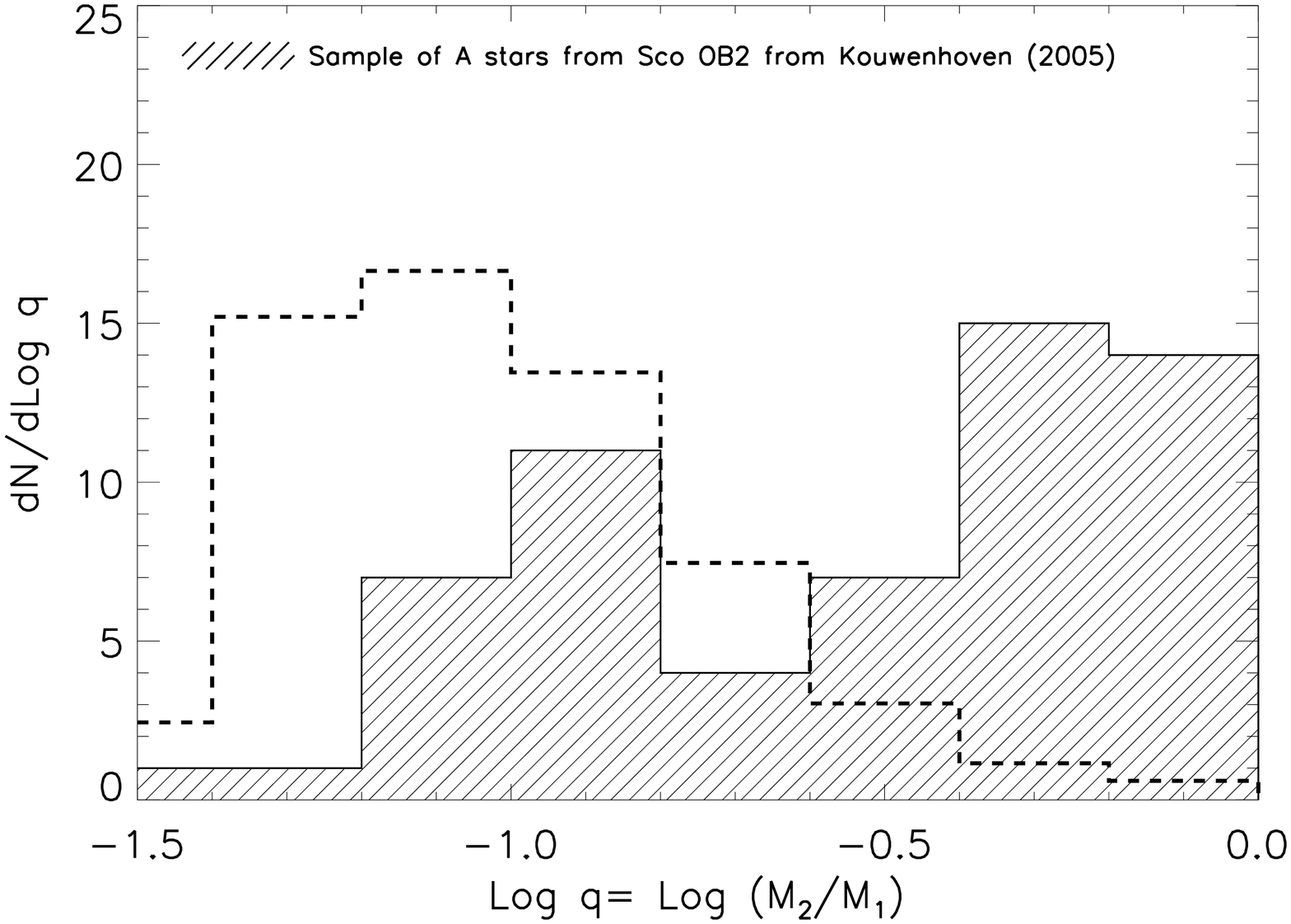}
\caption{\small{\textit{Companion mass ratio distributions and the IMF in the field.} From top to bottom the comparison between the observed CMRD and the log-normal field IMF is shown for M dwarfs, G stars in the field and for the sample of intermediate mass stars in Sco OB2 respectively. The hatched histogram represents the observed CMRD for the respective dataset of binary systems (see Section \ref{Datasets}). Superimposed with a dashed line is the CMRD generated for the same number of objects through random pairing from Bo2010. The KS probabilities are summarized in Table~\ref{table:KS test}.}}\label{cmr_field}
\end{figure}

The top panel of Fig.\ref{cmr_field} shows the CMRD for the sample of 27 M dwarf primary binary systems from \citet{Fischer1992}. The hatched histogram represents the observed distribution of $q$, while the dashed line is the CMRD generated by random pairing through Monte Carlo simulations from Bo2010 for the same range of mass ratios ($q\ge$0.2). The KS test gives a probability of $\sim$ 1\% that the observations are consistent with the IMF in the separation range 1-2400 AU.  From the figure it appears that there is an overabundance of equal mass binaries in the observed sample compared to the predictions of random pairing from the IMF.

In Fig.\ref{cmr_field} (central and bottom panels) we also present the results obtained for the other two datasets. In both cases we compare the observations (hatched histograms) with the simulated CMRDs (dashed line histograms) over the same range of $q$. On the basis of the KS test for the samples of solar type from the field and A type stars from Sco OB2 we get probability of $10^{-3}$ and $10^{-13}$ respectively that the CMRD is consistent with the field IMF in the separation range $\sim$30-1600 AU. The results are summarized in Table~\ref{table:KS test}. We find again that the observed mass ratios are more strongly peaked towards unity than in the simulations from the IMF.

This overall result suggests that we can reject the hypothesis that the CMRD is consistent with having been drawn from the IMF over the separation range $\sim$30-1600 AU and this statement seems to hold true independent of the primary mass and angular separation. In fact we checked with the KS test whether for each sample we see variations in the CMRD as a function of the angular separation.
Practically, for each one of the datasets we have considered different values of the angular separation within the range covered by the observations, and we evaluated for each of these separations the KS test probability of the CMRD inside this value being consistent with the CMRD outside. For any given separation we find probabilities less than 0.1\%. Therefore we did not see any evidence for dependence on orbital separation in any of the samples under study.

\begin{deluxetable*}{cccccc}
\tablewidth{0pt}
\tablecaption{KS test probabilities \label{table:KS test}}
\tablehead{\colhead{Sample - Type} &\colhead{Ref}\tablenotemark{a} &\colhead{No. Systems} & \colhead{Bo2010 (\%)} & \colhead{Flat CMF (\%)} & \colhead{ $dN/dM_{2}\propto M_{2}^{-0.4}$ (\%)}}
\startdata
Field - M & 1 & 27 & 1  & 58 & 24\\
Field - F/G & 2 & 30 &  $10^{-3}$ &  2 & 58  \\
ScoOB2 - A/late-B & 3 & 60 & $10^{-13}$ &  0.4 & 30\\
Pleiades - F/G  & 4 & 24 & $10^{-4}$ & 34  & 17 \\
$\alpha$ Persei - F/G & 5 & 18 & 0.1 &  27 &  89\\
Chamaeleon I - K/G & 6 & 13 & 17 &  30 &  76\\
Taurus- K/G & 7 & 40 & $10^{-11}$ &  45 &  2\\
\enddata
\tablenotetext{a}{References: (1) \citet{Fischer1992} , (2) \citet{Metchev2009}, (3) \citet{Kouwenhoven2005}, (4) \citet{Bouvier1997}, (5) \citet{Patience2002}, (6) \citet{Lafreniere2008}, (7) \citet{Kraus2011}.}
\end{deluxetable*}

We also note that the CMRDs generated for different primary masses through Monte Carlo simulations and shown in Fig. \ref{cmr_field} are significantly different. This is expected since the random pairing from the IMF predict a strong dependence of the CMRD on the primary mass. For example, a CMRD drawn from the IMF for primary mass near 1 $M_{\odot}$ should exhibit a peak near $q$=0.18, whereas for primary masses near 0.2 $M_{\odot}$ should decrease monotonically, as shown in Fig. \ref{cmr_field}.

As we do not see variation of the CMRD with angular separation, we can compare with the KS test also the observed CMRDs for these three datasets over the common range of $q$. The distributions of mass ratios for M dwarfs and solar type stars are consistent at the 6\% level. On the other hand the probability of the sample of intermediate mass stars from ScoOB2 being consistent with G/K stars and M dwarfs in the field is 36\% and 53\% respectively. These results suggest that we cannot reject the hypothesis that they are all drawn from the same parent distribution. 

\subsection{Results from clusters and associations} \label{Results_young}
\begin{figure}
\includegraphics[scale=.35]{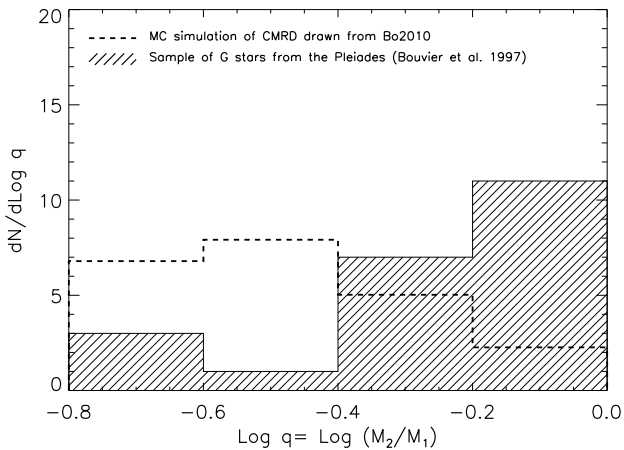}
\caption{\small{\textit{Companion mass ratio distribution for solar-type stars in the Pleiades.} The image shows the comparison between the observed CMRD in the Pleiades and the predictions from the IMF. We adopt the same legend as in figure \ref{cmr_field}. The probability from the KS test that observations are consistent with the IMF is less than 1\%.}}\label{cmr_G_Pleiades_aPer_Chamaeleon}
\end{figure}

We also compared the datasets of solar type stars from the Pleiades, $\alpha$ Per, Chamaeleon I and Taurus with the results of the MC simulations of the IMF over the range of $q$ spanned by each sample. In Figure \ref{cmr_G_Pleiades_aPer_Chamaeleon} we show the CMRD for the Pleiades.  The KS test probabilities that the observed CMRDs of all samples are consistent with having been drawn from the IMF are given in Table~\ref{table:KS test}. With regard to the Pleiades, $\alpha$ Per and Taurus (KS test probabilities of $10^{-4}$\%, 0.1\% and $10^{-11}$\%, respectively) we can reject the hypothesis of random pairing, while the higher probability (17\%) between the IMF and CMRD in Chamaeleon I does not allow us to rule out the null hypothesis in this case.
However, the number of objects (13) in Chamaeleon I sample is quite low. As we have shown in Section \ref{KS_test} the KS test can only distinguish extreme differences in distributions from such small samples. 

We should also note that in Taurus the IMF is peaked toward higher masses with respect to the field IMF \citep{Luhman2009}. The use of the proper mass distribution would bring the CMRD in Taurus in closer agreement with random pairing from the IMF. Hence we should be cautious in interpreting this preliminary result.

To summarize, at the moment in the Pleiades and $\alpha$ Per we can reject the possibility of the CMRD being drawn from the IMF for orbital separation between 20 and 600 AU, whereas concerning much younger regions, we rule out the random pairing from the field IMF only in Taurus in the separation range 5-5000 AU. Data from larger and different samples are needed to better constrain the result as a function of age and environment.

\subsection{Different Companion Mass Functions} \label{Results_flat}
Using the same Monte Carlo method, we have tested also whether the observed CMRD as function of primary mass and environment is consistent with other analytic forms of the CMRD. First of all we have considered a linearly flat companion mass ratio distribution (see Section \ref{KS_test}) and second we have tested the companion mass distribution $dN/dM_{2}\propto M_{2}^{-0.4}$ suggested by MH09. In Table~\ref{table:KS test} we report the KS probabilities for each dataset. 

\begin{figure}
\includegraphics[scale=.35]{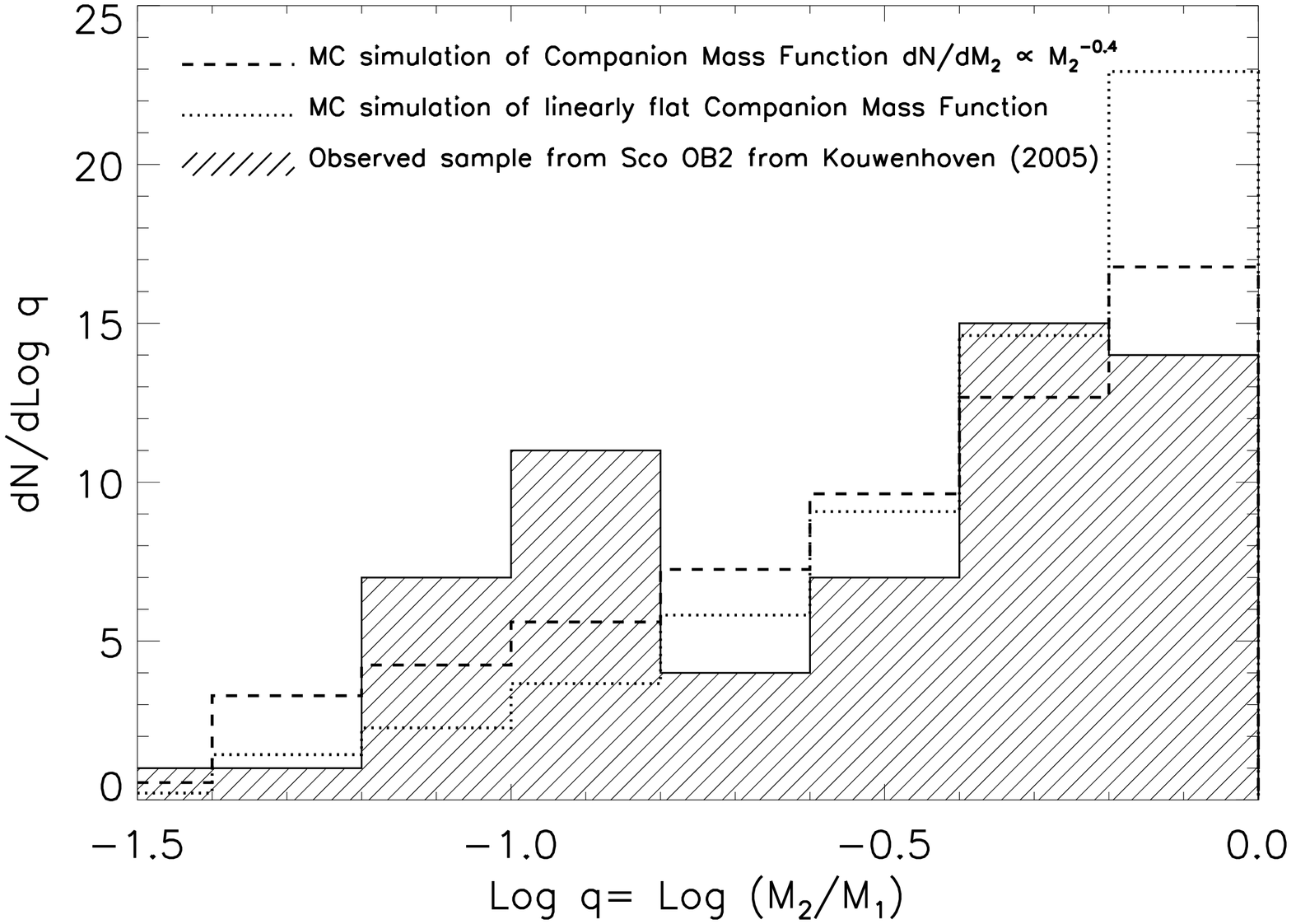}\\
\includegraphics[scale=.35]{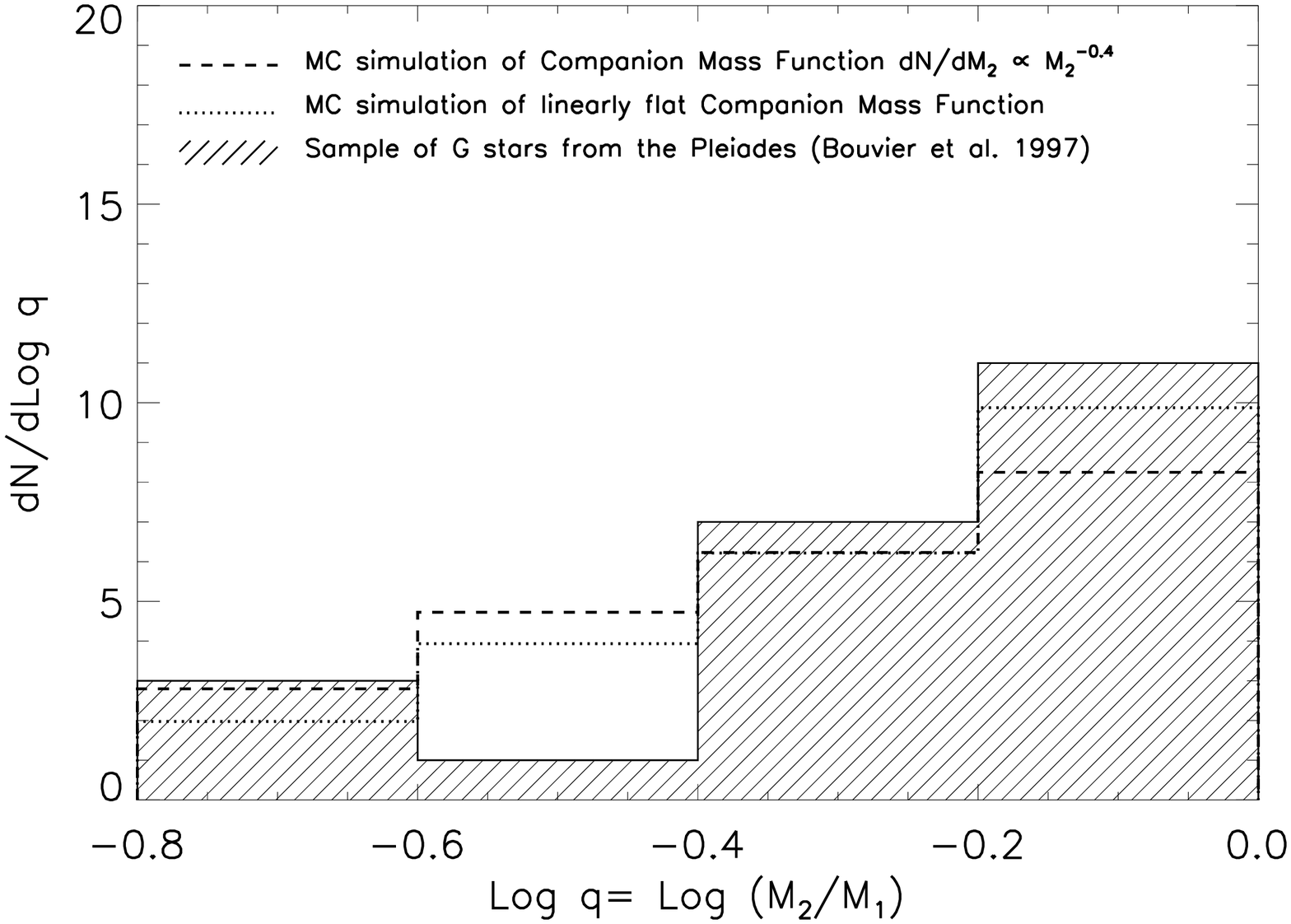}
\caption{\small{\textit{Test of other CMFs} Top: the figure shows the comparison between the observed CMRD for intermediate mass stars in ScoOB2 with a flat CMRD (dotted line) and a companion mass function of the form $dN/dM_{2}\propto M_{2}^{-0.4}$ (dashed line). We found a probability of less then 1\% that the observations are consistent with the flat CMF while a 30\% level of agreement with the CMF by MH09. Bottom: Comparison for the observed CMRD for solar-type stars in the Pleiades with the two choices of CMF. The KS test probabilities we obtained are 34\% and 17\% for the flat CMRD and MH09 CMF, respectively. The probabilities for all the other samples we have considered are given in Table~\ref{table:KS test}.}}\label{cmr_field_flat}
\end{figure}

Concerning the flat distribution, only for the sample of A and late B-type primary binaries in Sco OB2 we can reject the hypothesis that the two distributions are consistent. The comparison in this case is shown in the top panel of Figure~\ref{cmr_field_flat}. Note that the ScoOB2 dataset is the largest sample, placing the strongest constraints on possible differences. The MC simulations of a flat CMF for the young regions match well the observations (see e.g. bottom panel of Figure~\ref{cmr_field_flat}).
Regarding the CMRD provided by MH09 we find a KS probability exceeding 15\% for all samples (see  Table~\ref{table:KS test}) except for Taurus (2\%). 

We should keep in mind that the KS test is not suited to evaluate which is the best fit distribution. If we take as an example the results for the FM92 sample, the difference in the probability from 58\% to 24\% between the flat and M09 CMF in the context of the KS test does not have any significance.  Furthermore, the sample size of our datasets in the majority of cases prevents us from discriminating between log-normal, flat or other distributions (see Section~\ref{KS_test}). For this reason we have utilized a chi-square procedure to determine the best fit for the CMRD for a combined sample including all primary masses. 

\subsection{Chi-square best fit} \label{Chi-square best fit}
\begin{figure}
\includegraphics[scale=.35]{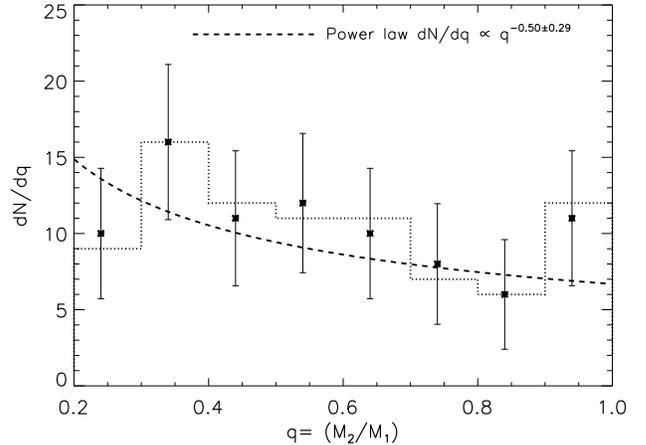}\\
\caption{\small{\textit{Chi-square best fit} Mass ratio distribution for the sample of primaries in the field with masses between 0.25-6.5 M$_{\odot}$ over the separation range 1-2400 AU. The best chi-square fit is a power-law $dN/dq \propto q^{\beta}$, with $\beta=-0.50\pm0.29$ ($\chi^2$= 0.7 with 7 degrees of freedom).}}\label{chi_fit}
\end{figure}

Motivated by the fact that the CMRD appears to be independent of angular separation over the range we are considering and that the distributions are not distinguishable, we combined together, over the common range of mass ratios ($q$=0.2-1), the samples of M dwarfs and G stars in the field and intermediate mass stars in ScoOB2 even though the separation ranges vary across the samples.
We then used the composite $q$ distribution to find the best fit. 
According to the chi-square test for $M_1$=0.25-6.5 M$_{\odot}$ the total mass ratio distribution follows a power-law $dN/dq \propto q^{\beta}$, with $\beta$=-0.50$\pm$0.29 ($\chi^2$= 0.7 with 7 degrees of freedom; see Figure \ref{chi_fit}).

This result is also in agreement with the value of $\beta$=-0.50 (45-900 AU) for B star primaries \citep{Shatsky2002} or with $\beta$=-0.4 for K dwarfs primaries \citep{Mazeh2003} in the orbital range 0-4 AU. \cite{Metchev2009} and \cite{Kouwenhoven2005}, already included in the sample under discussion, found $\beta$=-0.39$\pm$0.36 and $\beta$=-0.33 respectively.


\section{Discussion} \label{Discussion}
The results from the field and Sco OB2 described in section \ref{Results_field} show an overall trend of CMRDs more peaked towards equal mass values than predicted by random pairing from Bo2010. This result suggests that the capture hypothesis, at least for primary stars in the mass range 0.25-7 M$_{\odot}$  and orbital separation range 1-2400 AU, is not the major mechanism for binary formation, as it has been already proposed by a large number of previous studies \citep[e.g.][]{Clarke1991,Boffin1998,Bate2003}. 

Our findings appear to be in agreement with predictions from fragmentation theories of binary formation, even though we cannot discriminate between different fragmentation mechanisms.
In general, near-equal mass binaries are the most likely outcome of fragmentation in hydrodynamical simulations \citep[e.g.][]{Bate2000,Kouwenhoven2009}.
In these simulations, companions are expected to form by the fragmentation of massive accretion regions around stars in very early phases of star formation \citep{Goodwin2007} and to gain mass from the gas reservoir around them. In this process, even though a secondary star forms with an initial mass close to the opacity limit for fragmentation, it will accrete from the circumstellar material, reaching a mass roughly similar to the primary \citep{Kouwenhoven2009}.
These calculations generally predict a relation between mass ratio and separation, showing closer binaries with higher mass ratios than wider systems \citep{Bate2000,Bate2009}. This outcome differs from the observational evidence of a CMRD which is independent of separation from few to few thousands AU and suggests that some key element is still missing in our models of multiple formation.

It should be also noted that the field is likely a mixture of systems coming from very different environments \citep{Goodwin2010}. Binaries might have been processed in different ways \citep{Parker2009} and diverse star-forming regions may contribute in different degrees to the field.
Recent results from N-body simulations \citep{Parker2010} show that the CMRD for very low mass binaries is independent of dynamical evolution.  If this preliminary evidence holds for a broader range of primary masses, we can rule out the hypothesis that the CMRD was drawn from the IMF at any evolutionary stage, suggesting the current CMRD corresponds to the birth mass ratio distribution. If this is the case variations in the CMRD can be used to trace how different star formation regions contribute to the field. 

\begin{figure}
\includegraphics[scale=.35]{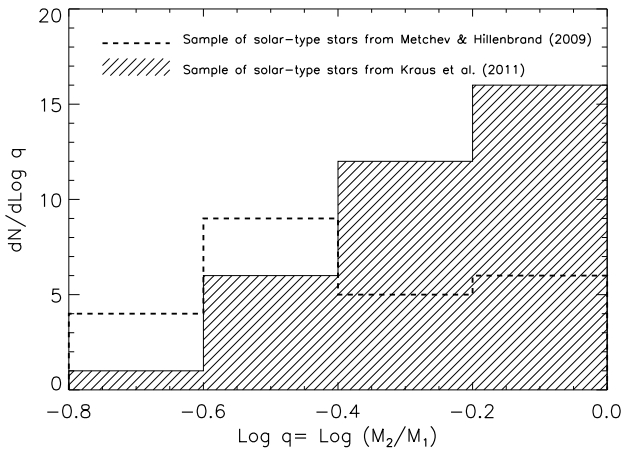}\\
\caption{\small{\textit{Companion mass ratio distribution for solar-type stars in Taurus \citep{Kraus2011} and in the field (MH09).} The image shows the comparison between the observed CMRD in Taurus for the sample of solar type stars primaries and the one observed in the field over the common range of $q$.}}\label{taurus_M09}
\end{figure}

Interestingly, if we compare the CMRD for solar-type primaries in the Pleiades and in the field (MH09) we obtain a probability of $\sim$1\%. Likewise (see also Figure \ref{taurus_M09}), we find a probability of $\sim$1\% with the KS test between the CMRDs for solar-type primaries in Taurus and the field (MH09). Perhaps bound open clusters like the Pleiades or extremely low density Taurus-like SFs do not contribute significantly to the field stellar population \citep[e.g.][]{Kroupa1995,Portegies2009,Adams2010}. This points toward using binary properties to understand which star formation events contribute most to the field.
Furthermore the comparison between the mass ratio distribution in Taurus and Pleiades return a 37\% KS probability, suggesting they are drawn from the same parent population. However the reason why the CMRD turns out to be similar in so different environments remains an open question.


\section{Conclusions} \label{Conclusions}
We have explored the connection between the CMRD and the IMF as a function of primary mass in the field and in a few examples of clusters and low density associations through Monte Carlo simulations. Using the KS test we determined the probability that the two distributions are consistent. We have also examined the probabilities that observed samples are consistent with having been drawn from a linearly flat mass ratio distribution and a CMF of the form $dN/dM_{2}\propto M_{2}^{-0.4}$. Finally we have found the best chi-square fit for a composite CMRD in the primary mass range 0.25-6.5 M$_{\odot}$ ($q\ge$0.2).

Our main results can be summarized as follows:
\begin{itemize}
\item [-] We can reject the hypothesis that the CMRD was drawn from the single object IMF for solar type stars (28-1590 AU and $q>$	0.1) and M dwarfs (1-2400 AU and $q>$0.2)  in the field and A and late-B type stars in Sco OB2 (29-1612 AU and $q>$0.05).
The observed CMRDs show a larger number of equal-mass systems than would be predicted by the IMF. This is in agreement with fragmentation theories of binary formation.
\item [-] We do not see evidence for variation of the CMRD of each sample (M and G stars in the field and AB stars in Sco OB2 association) with orbital separation in the ranges explored by the observations.
\item [-] Concerning the observed CMRDs for M dwarfs and G stars in the field, we obtain a probability of 6\% that they are consistent with each other over the same range of mass ratios. The CMRD for the sample of A and late-B type primaries in Sco OB2 is consistent with both the CMRDs of M and G stars. In other words, they all appear to be consistent with each other.
\item[-] Regarding the combined CMRD of M and G primaries in the field and intermediate mass primary binaries in Sco OB2 discussed above over the primary mass range 0.25-6.5 M$_{\odot}$, we obtain a chi-square best fit following a power law $dN/dq \propto q^{\beta}$, with $\beta=-0.50\pm0.29$, consistent with previous studies.
\end{itemize}
Certainly further binary studies in young clusters are needed to study the dependence of the CMRD on dynamical processes and to test possible variations in the mass ratio distribution as tracers of different star formation mechanisms.\\

\acknowledgments
We thank an anonymous referee for valuable suggestions. We are grateful to Richard Parker, Simon Goodwin, Cathie Clarke and Hans Zinnecker for the insightful discussions and the helpful comments. We are also thankful for the generous support of the Swiss National Science  Foundation through the grant  200021-132767.


\clearpage

\end{document}